\documentstyle[aps]{revtex}\tighten
\def \kkr {\rho^{\rm kk}}
\def \kkp {\pi^{\rm kk}_{\mu\nu}}

\begin{document}
\twocolumn[\hsize\textwidth\columnwidth\hsize\csname
@twocolumnfalse\endcsname

\title{Kaluza-Klein anisotropy in the CMB}

\author{John D. Barrow$^*$ and Roy Maartens$^\dag$}

\address{~}
\address{$^*$DAMTP, Centre for Mathematical Sciences,
Cambridge University, Cambridge~CB3~0WA, Britain}
\address{$^\dag$Relativity and Cosmology Group,
School of Computer Science and Mathematics, Portsmouth University,
Portsmouth~PO1~2EG, Britain}

\maketitle

\begin{abstract}

We show that 5-dimensional Kaluza-Klein graviton stresses can slow
the decay of shear anisotropy on the brane to observable levels,
and we use cosmic microwave background anisotropies to place
limits on the initial anisotropy induced by these stresses. An
initial shear to Hubble distortion of only $\sim
10^{-3}\Omega_0h_0^2$ at the 5D Planck time would allow the
observed large-angle CMB signal to be a relic mainly of KK tidal
effects.

\end{abstract}

 \vskip2pc]

Recent developments in string theory have inspired the
construction of braneworld models, in which standard-model fields
are confined to our 3-brane universe, while gravity propagates in
all the spatial dimensions. A simple 5D class of such models
allows for a non-compact extra dimension via a novel mechanism for
localization of gravity around the brane at low energies. This
mechanism is the warping of the metric by a negative 5D
cosmological constant~\cite{rs}. These models have been
generalized to admit cosmological branes~\cite{sms}, and they
provide an interesting arena in which to impose cosmological tests
on extra-dimensional generalizations of Einstein's
theory~\cite{m,mwbh,lmsw,l,m2}.

Modifications to general relativity in the direction of a quantum
gravity theory need to be consistent with increasingly detailed
cosmological observations. The premier cosmological test is
provided by cosmic microwave background (CMB) anisotropies. A
detailed calculation of CMB anisotropies predicted by braneworld
models is complicated by the need to solve the full 5D
perturbation equations, involving partial differential equations
for the Fourier modes. Up to now, only qualitative or special
results are known~\cite{m,mwbh,lmsw,l,m2}; in particular, the
Sachs-Wolfe effect has not yet been calculated because of 5D
effects~\cite{lmsw}. The 5D effects are carried by so-called
Kaluza-Klein (KK) massive modes of the graviton, which can
generate anisotropy on the brane. In view of the great complexity
of the full 5D problem, it is worth exploring partial aspects of
the problem. In this spirit, we impose some physically reasonable
assumptions on the 5D KK effects in order to estimate the CMB
large-angle anisotropy. Using COBE observations, this then leads
to constraints on the KK anisotropy. We find that the CMB imposes
significant limits on the initial anisotropy. It is even possible
that the observed large-angle anisotropy derives from anisotropic
KK gravitational effects.

Extra-dimensional modifications to Einstein's equations on the
brane may be consolidated into an effective total energy-momentum
tensor~\cite{sms,m}:
\begin{equation}
G_{\mu\nu}=\kappa^2 T^{\rm eff}_{\mu\nu}=\kappa^2
\left(T_{\mu\nu}+ T^{\rm loc}_{\mu\nu}+ T^{\rm kk}_{\mu\nu}
\right)\,. \label{6'}
\end{equation}
Since the brane cosmological constant $\Lambda$ is negligible at
the early times that we consider, we choose the bulk cosmological
constant so that $\Lambda=0$. The local effects of the bulk,
arising from the brane extrinsic curvature, are encoded in $T^{\rm
loc}_{\mu\nu} \sim (T_{\mu\nu})^2/\lambda$, and are significant at
high energies above the brane tension~\cite{mwbh}
\begin{equation}\label{he}
\rho\gtrsim\lambda\gtrsim 10^8~{\rm GeV}^4\,.
\end{equation}
The nonlocal bulk effects, arising from tidal stresses imprinted
on the brane by the bulk Weyl tensor~\cite{sms,m}, are the KK
modes carried by $T^{\rm kk}_{\mu\nu}$. We are interested here in
the astrophysically relevant case of small anisotropy and
inhomogeneity, and we adopt a long-wavelength velocity-dominated
approximation, so that inhomogeneous scalars vary slowly with
position and time-derivatives dominate over spatial derivatives.
In this approximation, we neglect the acceleration and vorticity
of the comoving 4-velocity $u^\mu$. Furthermore, since we are
interested in how anisotropic stresses from bulk gravitons affect
the shear anisotropy $\sigma_{\mu\nu}$ on the brane, we neglect
the matter and KK energy fluxes, and the tracefree anisotropic
matter stress. The effective total energy density, pressure and
anisotropic stress are therefore~\cite{m}
\begin{eqnarray}
\rho^{\rm eff} &=& \rho\left(1+\rho/ 2\lambda+ \kkr/
\rho\right)\,, \label{a'}\\ p^{\rm eff} &=& p\left(1+ \rho/
\lambda\right)+\rho\left( \rho/ 2\lambda+\kkr/ 3\rho \right)\,,
\label{b'}\\ \pi^{\rm eff}_{\mu\nu} &=& \kkp\,,~~\pi^{\rm
kk}{}_\mu{}^\mu=0\,.\label{d'}
\end{eqnarray}

The brane energy-momentum tensor separately satisfies the
conservation equations, $\nabla^\nu T_{\mu\nu}=0 $, and the
Bianchi identities on the brane imply that the effective
energy-momentum tensor is also conserved: $\nabla^\nu T^{\rm
eff}_{\mu\nu}=0 $. In general relativity, anisotropic stresses
slow the decay of shear anisotropy~\cite{b}. Without anisotropic
stress, this slower decay is impossible in general relativity, but
in the braneworld, KK graviton stresses can play a similar role to
anisotropic stress, as we show below. With our assumptions, the
conservation equations (see~\cite{m}) reduce to
\begin{eqnarray}
&&\dot{\rho}+\Theta(\rho+p) =0\,,\label{pc1}\\  && \dot{\rho}^{\rm
kk}+{\textstyle{4\over3}}\Theta \kkr+\sigma^{\mu\nu}\kkp =0 \,,
\label{lc1'}
\end{eqnarray}
where $\kkp$ is transverse as well as tracefree, a dot denotes a
comoving time derivative, and $\Theta=3\dot{a}/a$, where $a$ is an
average scale factor.

We also assume that the spatial curvature may be neglected. This
leads via the Gauss-Codazzi equations on the brane~\cite{svf,m2}
to a shear propagation equation and a Friedmann-like equation:
\begin{eqnarray}
\dot{\sigma}_{\mu\nu}+\Theta\sigma_{\mu\nu} &=& \kkp\,,\label{g1}
\\-{\textstyle{2\over3}}\Theta^2 +\sigma^{\mu\nu}\sigma_{\mu\nu}
+2\kappa^2\rho &=&-\kappa^2\rho^2/\lambda -2\kkr\,.\label{g2}
\end{eqnarray}

In Eqs.~(\ref{pc1})--(\ref{g2}), there is no evolution equation
for the nonlocal KK anisotropic stress $\kkp$. This is the
anisotropic stress imprinted on the brane by the 5D Weyl tensor,
and this nonlocal field cannot be predicted by brane-bound
observers since it includes 5D gravitational wave modes. The 5D
field equations must be solved to determine the brane dynamics
completely, and this also involves choosing boundary conditions in
the bulk. On the other hand, starting from a brane-bound
viewpoint, any choice of $\kkp$ that is consistent with the brane
equations, will correspond to a bulk geometry, which can be
locally determined in principle by numerical integration (or
approximately, close to the brane, by Taylor expanding the
Lie-derivative bulk equations given in~\cite{sms}). However,
numerical integration is very complicated (see~\cite{crss} for the
black hole case). Even if it can be successfully performed, it
will not give the global properties of the bulk. The bulk geometry
that arises for a given $\kkp$ may have unphysical boundary
conditions or singularities (e.g., the bulk corresponding to a
Schwarzschild black hole, with $\kkr=\kkp=0$, has a string-like
singularity and a singular Cauchy horizon~\cite{chr}).

We have no exact bulk solutions to guide us in a study of
cosmological anisotropy. The only relevant known
solution~\cite{bdel,sads} is the Schwarzschild-anti de Sitter bulk
that contains a (moving) Friedmann brane, which is the exactly
isotropic and homogeneous limit of our case, with $\kkp=0$. In the
absence of exact or numerical 5D solutions, we are forced to make
assumptions about the KK anisotropic stress $\kkp$ in order to
estimate its impact on the shear anisotropy of the brane. These
assumptions should be consistent with the brane equations above,
and physically reasonable, and are discussed below.

Observational constraints on the KK stresses arise from big bang
nucleosynthesis and from COBE measurements of large-angle CMB
temperature anisotropies. The KK energy density on the brane
introduces a new radiative degree of freedom at nucleosynthesis.
Assuming a maximum of 0.3 of a 2-component neutrino species, and
helium limits of 0.228 to 0.248, this gives
(compare~\cite{bdel,lmsw})
\begin{equation}\label{ns}
\left| \kkr/ \rho\right|_{\rm ns} \lesssim .024\,.
\end{equation}
The large-angle CMB anisotropies are constrained by
\begin{equation}\label{cmb}
{\Delta T\over T}\sim s(t_{\rm ls}) \lesssim 10^{-5}\,,~
s=\sqrt{s_{\mu\nu}s^{\mu\nu}}\,,~~s_{\mu\nu}={\sigma_{\mu\nu}\over
\Theta}\,,
\end{equation}
where $t_{\rm ls}$ is the time of last scattering. By
Eq.~(\ref{g2}),
\begin{equation}\label{s}
s^2={\textstyle{2\over3}}-2\kappa^2\rho \left[1+{\rho/ 2\lambda} +
{\kkr/\rho}\right]/\Theta^2\,,
\end{equation}
and then Eq.~(\ref{ns}) gives the nucleosynthesis limit
\begin{equation}\label{ns'}
s(t_{\rm ns})\lesssim 0.13\,.
\end{equation}

For small anisotropy, the volume expansion of the universe is
determined by the isotropic matter source [we assume that $\kkr=0$
in the background isotropic solution of Eq.~(\ref{g2})]. This
amounts to treating $a$ to lowest order as the scale factor for
the isotropic Friedmann braneworld. In this approximation, and
with equation of state $p=(\gamma-1)\rho$, with $\gamma$ constant,
Eqs.~(\ref{pc1}) and (\ref{g2}) lead to
\begin{eqnarray}
&&\Theta =(2t+\beta)/[\gamma t(t+\beta)]\,,~ \rho=
4/[3\kappa^2\gamma ^2t(t+\beta)]\,,\label{a}\\ &&\beta=
\sqrt{8/\lambda}/\kappa \lesssim 10^{-9}~\mbox{sec}\,,
\end{eqnarray}
where the bound follows from Eq.~(\ref{he}). This is in agreement
with the Friedmann brane solutions given in~\cite{bdel}. (If we
take into account the nonlocal energy density in the background
solution, i.e. $\kkr \neq0$, we can generalize this solution when
$\gamma={4\over3}$: $\Theta$ is still given by Eq.~(\ref{a}), with
$\beta$ replaced by
$\tilde{\beta}=\beta[1+2\kkr/(\kappa^4\lambda\rho)]$, while
$\rho\to (\tilde{\beta}/\beta)\rho$.) The low-energy limit is
$\beta\to0$, when the general relativity solutions are regained.

Cosmological anisotropy on the brane has been considered in recent
papers. A qualitative description of the role of nonlocal
anisotropic stress in cosmology is given in~\cite{m,m2}, and
perturbative analysis on large scales is developed in~\cite{lmsw},
but without finding or assuming a form for $\kkp$.
In~\cite{mss,chm}, Bianchi~I dynamics on the brane, with vanishing
spatial curvature, is studied. In particular, the local
extra-dimensional modifications to general relativity introduce a
novel feature to early-time dynamics~\cite{mss}: instead of shear
domination, there is {\em matter} domination at early times, and
the relative shear anisotropy $s$ is a maximum when
$\rho/\lambda=(2-\gamma)/( \gamma-1)$.

Here we investigate the evolution of shear anisotropy in an
inhomogeneous universe, at early times (when $\Lambda$ may be
neglected) and in the absence of anisotropic stress from fields on
the brane, i.e. $\pi_{\mu\nu}=0$. We need to determine the effect
of KK anisotropic stress $\kkp$ on braneworld shear. In principle,
the 5D Weyl tensor is determined by the solution of the 5D field
equations, and its projection onto the brane then determines
$\kkp$. In this way, the 5D solution determines an effective
evolution equation on the brane for $\kkp$. In practice, we know
of no way to find or estimate this 5D determination of $\kkp$. We
emphasize that $\kkp$ is a 4D quantity, which is an effective 4D
anisotropic stress, even though its evolution is governed by 5D
graviton dynamics. Thus it seems reasonable as a first
approximation to assume that $\kkp$ behaves qualitatively like a
general 4D anisotropic stress. This should be general enough to
cover a wide range of bulk graviton effects. Then we can estimate
the KK effect on large-angle CMB anisotropies. (Qualitative
estimates in the absence of 5D solutions have also been used to
estimate CMB anisotropies in different braneworld models, where
gravity is modified at large scales rather than high
energies~\cite{bs}.)

There is an ansatz that includes all examples of anisotropic
stresses studied in relativistic cosmologies and is physically
well motivated. According to this ansatz, in the large-scale
velocity-dominated approximation, the time evolution of tracefree
anisotropic stress is proportional to the energy density of the
anisotropic source. This general form includes the known cases of
collisionless radiation, (4D) gravitational waves, electric and
magnetic fields, strings and walls~\cite{b}. We assume that the KK
anisotropic stress on the brane behaves qualitatively in a similar
way, i.e.,
\begin{equation}\label{as}
\kkp  =\kkr  C_{\mu\nu}\,,~~~ \dot{C}_{\mu\nu}=0\,,
\end{equation}
where $\sqrt{C_{\mu\nu}C^{\mu\nu}}$ is $O(1)$, while $\kkr$ is
perturbatively small relative to the matter energy density $\rho$,
as shown by Eq.~(\ref{ns}). The energy conservation
equation~(\ref{lc1'}) and the shear evolution equation~(\ref{g2})
imply that
\begin{eqnarray}
\dot{r}&=&(\gamma-{\textstyle{4\over3}}) \Theta r
-\sigma^{\mu\nu}C_{\mu\nu}r\,,~~r\equiv{\kkr/ \rho} \,,\label{b}\\
\dot{s}_{\mu\nu} &=& -(\Theta+ {\dot{\Theta}/\Theta})s_{\mu\nu}+
\left({\kappa^2\rho/ 3H}\right) rC_{\mu\nu}\,.\label{c}
\end{eqnarray}
It is important to notice the special situation that arises in
Eq.~(\ref{b}) when the perfect fluid background is radiation
($\gamma ={4\over3}$). In this case the stability of the isotropic
solution is determined at {\em second order}. Linearization about
the isotropic expansion would lead to a single zero eigenvalue
associated with the shear eigenvalue.

Consider first the simpler case when $\gamma <\frac 43$. The
second term on the right-hand side of Eq.~(\ref{b}) can be
neglected with respect to the first and the evolution of the KK
energy density is
\begin{equation}
\kkr=N(\vec x)[t(t+\beta)]^{-4/3\gamma}\,. \label{e}
\end{equation}
Although this analysis appears to hold for the models with $\gamma
>\frac 43, $ it does not. In these cases $\dot{r}>0$, and the
KK anisotropic stresses would have a gravitational effect that
grows with time, invalidating the assumption of Eq.~(\ref{a}) that
the volume expansion is well approximated by that of the isotropic
solution. The solution for the shear is obtained from the solution
of Eqs.~(\ref{c}) and (\ref{e}):
\begin{eqnarray}
\sigma _{\mu \nu }&=& [t(t+\beta)]^{-1/ \gamma }\Bigl\{\Sigma
_{\mu \nu }(\vec x) + \nonumber\\ && {}+\kappa^2N(\vec x)C_{\mu
\nu }(\vec x)\int[t(t+\beta)]^{-1/3\gamma} dt\Bigr\}\,. \label{f'}
\end{eqnarray}
In the low energy regime at late times, $t\gg \beta$,
\begin{equation}\label{f}
\sigma_{\mu\nu}=\Sigma_{\mu\nu}t^{-2/\gamma} +
{3\gamma\kappa^2\over3\gamma-2}NC_{\mu\nu}t^{-(8-3\gamma)/3\gamma}\,.
\end{equation}
This has the same form as the general relativity result when
anisotropic matter stresses are present~\cite{b}. When the KK
anisotropic stress is absent, the solution is determined by the
rapidly falling $\Sigma_{\mu \nu }$ mode that is familiar from
studies of simple anisotropic Bianchi Type~I universes with
isotropic stresses. However, when KK anisotropic stresses are
present, so that $C_{\mu \nu }\neq 0$, these stresses slow the
decay of the shear because of the anisotropic effect of their
pressures. The shear evolution becomes increasingly dominated by
the $C_{\mu \nu }$ mode at late times. Note in particular that
during a dust-dominated ($\gamma =1$) era the two modes evolve as
$\sigma _{\mu \nu }=\Sigma _{\mu \nu }t^{-2}+ 3\kappa^2NC_{\mu\nu}
t^{-5/3}$. The KK mode dominates at large $t$ whenever
$\gamma>{2\over3}$.

Next, we consider the radiation-dominated solution ($\gamma
={4\over3}$). This is physically the most relevant but is
mathematically distinct. The variables $\Theta$ and $\rho$ take
their isotropic universe values, and Eqs.~(\ref{b}) and (\ref{c})
have a solution of relaxation form, with
$\dot{s}_{\mu\nu}\rightarrow 0$,
\begin{eqnarray}
\kappa^2\kkr&=&[C_{\alpha\beta}(\vec x) C^{\alpha\beta}(\vec x)\,
t(t+\tilde{\beta})]^{-1}\times \nonumber\\&&~~\times [ F(\vec x)+
\ln (4t^2+4{\beta}t-{\beta}^2)]^{-1}\,, \label{h}\\ \sigma_{\mu
\nu } &=& 4\kappa^2\kkr(t,\vec x)\left[
{t(t+{\beta})(2t+{\beta})\over 4t^2+4{\beta}t-{\beta}^2}\right]
C_{\mu\nu}(\vec x)\,. \label{i}
\end{eqnarray}
At low energies, $t\gg\beta$,
\begin{eqnarray}
\kappa^2\kkr &=& \left[2t^2\, C_{\alpha\beta} C^{\alpha\beta} \,
\ln( t/t_*)\right]^{-1}\,, \\ \sigma_{\mu \nu } &=& C_{\mu\nu}
\left[t\, C_{\alpha\beta} C^{\alpha\beta}\, \ln(
t/t_*)\right]^{-1}\,,~~t_*=t_*(\vec x)\,,
\end{eqnarray}
where $t_*$ is some spatially-varying initial time, with $t>t_*$.
This has the same form as the general relativity result for matter
anisotropic stress~\cite{b}. For a diagonal metric with expansion
scale factors $a_i(t)$, it leads to $a_i(t)\propto t^{1/2}(\ln
t)^{n_i}$, where the constants $n_i$ are of order unity, such that
$\Sigma n_i=0$, and are determined by the eigenvalues of the
symmetric tracefree matrix $C_{ij } $, which specifies the KK
anisotropy shape. For example, with an axisymmetric ($a_1=a_2\neq
a_3)$ shear anisotropy, we have $a_1(t)\propto a_2(t)\propto
t^{1/2}(\ln t)^{1/4}$ and $a_3(t)\propto t^{1/2}(\ln t)^{-1/2}$.

In the radiation case we note the distinctive slow evolution of
the shear anisotropy and non-additive perturbation to the scale
factors in the presence of KK anisotropic stress. By
Eq.~(\ref{i}), the ratio of shear to Hubble expansion evolves as
\begin{equation}
s\propto {t(t+{\beta}) \over [F(\vec x)+ \ln
(4t^2+4{\beta}t-{\beta}^2)](4t^2+4{\beta}t- {\beta}^2)}\,,
\end{equation}
so that at late times it falls only {\em logarithmically} in time:
\begin{equation}
s \rightarrow {1/[ 8\ln (t/t_*)]}\,,~~t\gg\beta>t_*\,.\label{log}
\end{equation}
In the absence of KK anisotropic stresses, $s$ would fall off as
$s \propto [t(t+\beta)]^{1/4}/(2t+\beta)\rightarrow t^{-1/2} $.
Even if $s$ were $O(1)$ close to the string or Planck scales, it
would have become observationally insignificant by the epoch of
primordial nucleosynthesis ($t_{\rm ns}\sim 1-100$~s) or last
scattering of the CMB ($t_{\rm ls}\sim 10^{13}$~s).

The inclusion of the bulk graviton anisotropic stress $\kkp$,
which has been largely neglected in earlier studies of anisotropic
braneworld evolution, completely changes the picture of long-term
evolution. This anisotropic stress completely determines the
evolution of small expansion anisotropies. Physically, the
isotropic stress tends to isotropize the expansion, while the
anisotropic stresses tend to resist this isotropizing effect.
Since to first order they are both radiation fluids, it is the
second-order effect of the KK anisotropy that dominates. Its
logarithmic influence reflects its second-order character. In
general relativity, this behaviour is only possible if there are
anisotropic stresses from matter fields or (4D) gravitational
waves. In the braneworld, we can get this behaviour even in the
absence of $\pi_{\mu\nu}$, because 5D graviton stresses imprinted
on the brane can play the role of $\pi_{\mu\nu}$.

The most interesting feature of the radiation-era solution is that
the slow decay of the shear anisotropy allows the shear distortion
to have potentially observable consequences. The most sensitive
effect is on the CMB rather than nucleosynthesis (this would {\em
not} be the case if the shear decay was a power of time rather
than logarithmic). If gravity in the bulk induces an initial
anisotropic stress on the brane, $ s(t_{\rm in})$, then during the
radiation-dominated era, $s\propto [\ln (t/t_*)]^{-1}$, with
$t_*(\vec x)$ time-independent, by Eq.~(\ref{log}). During the
short interval of dust-dominated evolution from the equal-density
epoch $t_{\rm eq}$ until $t_{\rm ls}$, we have $s \propto
t^{-2/3}$ by Eq.~(\ref{f}). Thus the CMB temperature anisotropy on
large angular scales will be
\begin{equation}
\frac{\Delta T}T\sim s(t_{\rm ls})\approx s(t_{\rm in})\left(
\frac{t_{\rm eq}}{t_{\rm ls}}\right)^{2/3}\left[ \frac{\ln (t_{\rm
in}/t_*)}{\ln (t_{\rm eq}/t_*)}\right]\,.  \label{j}
\end{equation}

Crucially, the magnitude of $\Delta T/T$ is only logarithmically
dependent on $t_{\rm in}$ and $t_*$. Using $1+z_{\rm ls}=1100$ and
$1+z_{\rm eq}=2.4\times 10^4\Omega _0h_0^2,$ where $\Omega _0$ is
the present total matter density of the universe in units of the
critical density and $h_0$ is the Hubble parameter today in units
of 100~km\,s$^{-1}$\,Mpc$^{-1}$, we find
\begin{equation}
\frac{\Delta T}T \sim  4.6\times 10^{-2}\Omega _0^{-1}h_0^{-2}\,
s(t_{\rm in}) \left[ \frac{\ln (t_{\rm in}/t_*)}{\ln (t_{\rm
eq}/t_*)}\right] \,.
\end{equation}
The time $t_{\rm in}$ could reasonably be taken as the 5D Planck
time $t_5$. By Eq.~(\ref{he}) the 5D Planck mass is subject to
$M_5\gtrsim 10^8$~GeV, so that $t_5\lesssim 10^{11}t_4$, where
$t_4\approx 10^{-43}$~sec is the 4D Planck time. For $t_*\sim
t_4\approx 10^{-43}$~sec and $t_{\rm in}\sim 10^{10}t_*$, the
logarithmic term would be $\sim {1\over5}$, and is relatively
insensitive to quite large changes in these quantities. Thus,
using Eq.~(\ref{cmb}),
\begin{equation}
s(t_{\rm in}) \lesssim10^{-3}\Omega _0h_0^{2}\,.
\end{equation}
This is a much tighter constraint on the initial anisotropy than
obtained from nucleosynthesis, Eq.~(\ref{ns'}). The observed
large-angle temperature anisotropy in the CMB may have been
contributed by bulk graviton effects in the very early universe if
they have an initial amplitude of $ \sim 10^{-3}\Omega _0h_0^2$.
This anisotropy level is too low to have an observable effect on
the output from primordial nucleosynthesis.

We have shown that the tidal stresses induced on a perfect-fluid
braneworld by bulk gravitons are the dominant factor in
determining the evolution of its anisotropic distortion. During
the radiation era this distortion falls only logarithmically in
time relative to the expansion rate of the brane and can
contribute a significant component to the large angular scale
anisotropy of the CMB. An initial $\sigma /\Theta $ ratio of only
$s(t_{{\rm in}})\sim 10^{-3}\Omega _0h_0^2$ would allow the
observed CMB signal to be a relic mainly of KK graviton tidal
effects.

JDB thanks Akihiro Ishibashi for discussions.

\end{document}